\documentclass{iopart}

\usepackage{epsfig}
\usepackage{floatflt}
\usepackage{wrapfig}
\usepackage{ulem}
\usepackage{layout}
\usepackage{fancyhdr}
\usepackage{extramarks}

\begin{document}
\title{Simple field theoretical approach of Coulomb systems. Entropic effects.}

\author{D. di Caprio \footnote{E-mail: dung.di\_caprio@upmc.fr} and J.P. Badiali}
\address{Laboratory of Electrochemistry and Analytical Chemistry,
University Paris 6, CNRS, ENSCP, B.P. 39,
4, Place Jussieu, \ 75252 Paris Cedex 05, France}
\author{M. Holovko}
\address{Institute for Condensed Matter Physics,
National Academy of Sciences\\1 Svientsitskii Str., 79011 Lviv, Ukraine\\}

\begin{abstract}
We discuss a new simple field theory approach of Coulomb systems.
Using a description in terms of fields, we introduce in a new way the
statistical degrees of freedom in relation with the quantum mechanics.
We show on a series of examples that these fundamental entropic effects can
help account for physical phenomena in relation with Coulomb systems
whether symmetric or also asymmetric in valence. On the overall, this
gives a new understanding of these systems.

\end{abstract}

\section{Introduction}
Recently, we have developped a simple field theory (FT) approach to describe
liquids \cite{ddcjpbjphysA} and more specifically electrolytes at interfaces.
The use of density fields as basic elements of description rather than particle
density distributions comes together with a radically different way of
accounting for the statistical degrees of freedom of the system.
It is the quantum mechanics which sets the correct combinatory of the
states by telling us which two states can be distinguished in the phase space.
In this paper, on a series of examples related to Coulomb systems, we illustrate
how within our approach the denombration of the states can help us interpret
and give an original insight into a series phenomena related to Coulomb
systems.
In the first part of the paper, we discuss general aspects and consequences
of the formalism.
In the second part, we illustrate the entropic role of the quantum mechanics in
Coulomb systems.

\section{A new and simple field theory to describe liquids, exact results}
The fundamental choice of our FT is to take directly the density of matter
$\rho(\mathbf{r})$ as the basic real valued field variable.
As the densities fluctuate, we then fix the chemical potential
$\mu$ and set the partition function to be the grand potential.
It is then a functional integral according to
$-\beta P V = \ln \Xi = \ln [\int \mathcal{D}\rho \; e^{-\beta H[\rho]}]$, where 
$\beta$ is the inverse temperature.
The Hamiltonian functional $H$ corresponds to a simple meaningful mean field
approximation for the chemical potential
$\beta\mu=\ln (\rho(\mathbf{r})\Lambda^3) + \int \rho(\mathbf{r}') v(|\mathbf{r}-\mathbf{r}'|)
        d\mathbf{r}' $ which leads to
$\beta H[\rho] =\int \rho(\mathbf{r})
    \left[ \ln (\rho(\mathbf{r})/ \bar{\rho}_z) -1   \right]
         d\mathbf{r} + \int \rho(\mathbf{r}) \rho(\mathbf{r}')
      v(|\mathbf{r}-\mathbf{r}'|)\,d \mathbf{r} d\mathbf{r}' $
where $\bar{\rho}_z=e^{\beta\mu}/\Lambda^3$ is the activity.
In \cite{ddcjpbjphysA}, we have shown that provided a renormalization, this
theory reproduces exactly the standard statistical mechanics.
The peculiarity of the approach lies in the first term which is entropic and accounts
for the quantum mechanics denombration of states with \textit{i})~the elementary volume in the phase space
$\Lambda^3$ and \textit{ii})~the principle of indiscernibility for particles both properties
which are, for $N$ particles, traditionally given by the
${\prod_1^N d \mathbf{r}_i}/{(N! \Lambda^3)}$ measure.

\textit{Dyson like relations or equations of movement.}
A direct consequence of the integral functional formalism, can be seen in the
invariance of the functional integral with respect to a change of the dummy
field variable, for any functional $A[\rho]$ we have for the averages
$   \left< \delta A[\rho]/\delta \rho\right> = 
   \left< A[\rho](\delta \beta H[\rho]/\delta \rho)\right>$.
Starting from the partition function, the application of this relation
gives a new expression of the chemical potential:
$\beta\mu = \langle \ln[ \rho(\mathbf{r}) \Lambda^3]\rangle +
  \beta \int v(|\mathbf{r}-\mathbf{r}'|)\langle \rho(\mathbf{r}') \rangle d\mathbf{r}'$.
This relation has been discussed in view of more standard expressions
\cite{Dyson}.
Where field theory shows a simple mean interaction potential term and where the
average of the entropic logarithmic term includes more intricate correlations,
in contrast,
standard expressions emphasize the interaction potential, the
logarithmic term being the trivial ideal system. 
This displacement in the way correlations are treated, naturally suggests new
approximations.

\textit{New perspectives on old exact relations.}
We have rederived in \cite{Dyson} the exact density contact theorem
and the Virial theorem. We have pointed out that our approach provides yet
another point of view on this relation emphasizing the role of the logarithmic
functional in the demonstration.

\section{Entropic effects in Coulomb systems.}
\textit{Coulomb systems are peculiar mixtures:} by this we mean that we have two
species bound by the condition of electroneutrality.
On the one hand, we have entropic properties related with densities of each of
the two indiscernable species. On the other hand, in Coulomb systems we like
to focus on the difference of the densities, the charge.
Our point is to treat on the same footing the purely entropic properties
and the one related to the potential as they appear at the same level in our
Hamiltonian. And to show that this will lead to strong correlations between the
different properties.
Intuitively, when we change the charge, we also modify the relative number of
the species therefore their purely entropic combinatorics contribution.
We hereafter discuss simple \textit{point ions} to focus on these entropic terms.

At a charged interface the mean field approximation gives the Gouy-Chapman theory.
Discussing fluctuations beyond the mean field theory correspond to an expansion of the 
entropic part of the Hamiltonian which for Coulomb systems is
 $ H^{ent} = \int d\mathbf{r} \;\{ \rho_+(\mathbf{r})\left[\ln (\rho_+(\mathbf{r})/\bar{\rho}_+)-1\right]
   + \rho_-(\mathbf{r})\left[\ln (\rho_-(\mathbf{r})/\bar{\rho}_-)-1\right]\}$
with separate terms corresponding to each ion. But
for the more natural variables for coulombic systems,
the charge $q=\rho_+-\rho_-$ and the total density $s=\rho_++\rho_-$,
the entropic functional mixes the fields
as we have $\rho_\pm=(s\pm q)/2$.
Note a peculiarity in FT is that linear combinations of the fluctuating fields
$q$, $s$ are natural and not simply combinations for the average quantities.
In the expansion, the quadratic part of the Hamiltonian includes the Coulomb
potential interaction.
With this quadratic term, we obtain the Debye limiting law in the bulk.
The specificity of the FT lies in the higher order terms which include products of $q$ and $s$ fields
   $\delta H^{ent} = -1/(2\bar{\rho}) \int q^2(\mathbf{r})s(\mathbf{r})\;d\mathbf{r}
   + ...$ in local coupling terms.

\subsection{Application to inhomogeneous systems.}

\textit{Ionic depletion at a neutral interface}\cite{ionicprofile}
For a neutral interface and a symmetric electrolyte, there is intuitively no
profile for the charge at a neutral interface. However, the loop expansion of
the theory shows a profile of the quadratic fluctuations of the charge
with a simple expression with the distance to the wall
and a dependance on {$\eta=K_D^3/(8\pi\rho)$} which
characterizes the Coulomb interaction strength where $K_D$ is the inverse
Debye length.
Physically, some fluctuations are frustrated at the vicinity of the interface
due to the absence of ions in the other half space.
The role of the entropic term is to imply that this coulombic effect
is coupled to the total density field although this quantity is not directly
related to Coulomb interaction.
The pertinence of this profile is further justified as it verifies the contact
theorem which is not the case of the intuitive Gouy-Chapman or MSA (Mean
Spherical Approximation) approaches.

\textit{The anomalous capacitance behaviour.}
Experimentally and in numerical simulations, we find systems where the electric
capacitance decreases at low reduced temperature. This effect is non intuitive
as one can expect the electric response of the system to decrease with increasing thermal
agitation which is what the Gouy-Chapman and MSA theories predict.
This phenomenon appears at low reduced temperature when the Coulomb interaction
becomes stronger.
Within the field theory, the phenomenon can easily be understood with the
previously described depletion in the density profiles which naturally accounts
for the decrease in the electric capacitance.
For Point ions \cite{anomalouszofia}, we have obtained
a simple analytic expression of the capacitance.
This has been corrected to account for the size of the ions \cite{densityprofile}.
The comparison with the numerical simulations is rather good and
the simplicity of the expressions and of the corrections indicate we capture
the significant physical effect.

\subsection{More entropic effects for multivalency}
Entropic effects are also important for valence asymmetric systems,
$z_+:z_-$ electrolytes where the electroneutrality condition has now modified the
number balance of the ions.

\textit{Anomalous capacitance for valence asymmetry.}
In \cite{anomalousasym}, we have considered capacitance curves
for different valencies, the curves are rather scattered but one finds
that one should first redefine the temperature $T^* \rightarrow T^*_s=T^*/(z_+z_-)$ which
is also a scaling with the ionic strength.
However, the ionic strength $z_+z_-$ scaling does not discriminate between a 2:2
and a 4:1 electrolyte.
The FT introduces a distinct parameter $z_{as}=(z_+-z_-)/\sqrt{z_+z_-}$ which is
really characteristic of the valence asymmetry.
As in simulation results, FT predicts an increase of the differential
capacitance peak with $ z_{as}$.
This parameter appears naturally in FT as we can perform simple operations
on the fluctuating fields and again it is related to the entropic term.

\textit{Potential of zero charge (PZC) for size and valence asymmetric systems.}
For the neutral interface, numerical simulations results show the existence
of a spontaneous polarisation due to the asymmetry in size and/or in valence \cite{henderson}.
The first is rather intuitive as the smaller ions by reaching
nearer the interface induce a polarisation.
Less intuitive is the polarisation due to valence asymmetry. Indeed, in figure 1,
standard approaches in liquids state physics \cite{henderson} do not account for
such effect.
\begin{figure}
\begin{center}
\hspace*{-10ex}\scalebox{0.20}{\includegraphics*{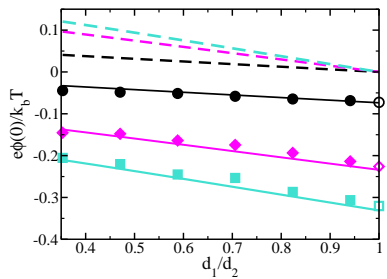}}
\end{center}
\caption{PZC for 3:1 electrolytes as a function of $d_1/d_2$ ionic diameters ratio.
The symbols are for the numerical simulation with the corresponding FT plus MSA
predictions (see text), concentrations are 0.033, 0.333, 0.666 Mol/l from top down.
The dashed line are for the MSA in the reverse order for the concentrations.}
\end{figure}
On the contrary, FT predicts the polarisation due to multivalency and we show
results obtained by adding the MSA theory for size asymmetry to the FT result
for point ions (size asymmetry being immaterial for point ions) \cite{vasym}.
The sign of the PZC is in agreement with the charge contact theorem \cite{chargecontact}
and with the simulation results. The simplicity of the correction
suggest again that we focus on the adequate physical effect.

\textit{Ionic criticality and valence asymmetry.}
We have derived in the FT an equation of state for ionic systems.
In the loop expansion, the pressure is a series in terms of $\eta$:
 $\beta P = {\rho}\left(1-\frac{1}{3}{\eta}-d_0{ z_{as}^2}{\eta}^2+...\right)$.
The linear term $-\eta/3$ is the standard Debye term,
further we have a simple dependence with the ${ z_{as}}$ parameter.
We have generalized this expression to account for hard sphere effects
by using the MSA pressure%
\footnote{Corresponds to replacing $K_D \rightarrow 2\Gamma$ 
the MSA screening distance parameter, $2\Gamma=\sqrt{1+2K_D}-1$}
and calculated the asymmetry contribution $d_0$ using the MSA correlation function.
The term in the pressure associated with valence asymmetric systems gives a
simple description of the decreasing critical temperature and increasing
critical density in agreement with Monte Carlo (MC) results \cite{Fisher}.
Respectively for the 1:1, 2:1, and 3:1 electrolytes
the critical temperatures and densities are
$T^*=0.079, 0.060, 0.038$ and $\rho^*= 0.0145, 0.021, 0.034$.
This seems to indicate that the physics is likely to be associated with the simple
number balance effect found in multivalent systems.

\section{Conclusion}
Within a FT framework, we have shown how the formalism can introduce a new point
of view on Coulomb systems.
The basic and fundamental quantum mechanical counting of states can lead to
important consequences in Coulomb systems which have at least two indiscernable
species.
These entropic effects lead to a unavoidable correlation between charge and
density fluctuations. This simple statement is sufficient to give
meaningful interpretations, we believe this is useful to complete the standard
liquid state approaches generally focused on excluded volume effects.\\

\bibliographystyle{iopart-num}

\end{document}